# AI, Expert or Peer?
# Provider Biases and Feedback Uptake among Pre-Service Teachers

Lucas Jasper Jacobsen[1], Ute Mertens[2], Thorben Jansen[3], Kira Elena Weber[4]

## Highlights

- Pre-service teachers received LLM, expert, and peer feedback
- LLM feedback was correctly identified close to random guessing
- Perceived source shaped feedback perceptions
- LLM feedback was perceived most positively when falsely ascribed to a human source
- Perceived feedback source influenced feedback perception but not feedback uptake

[1]Universität Hamburg, Sedanstraße 19, 20146 Hamburg, Germany, first author, corresponding author, ORCID: 0009-0001-6967-3751, lucas.jacobsen@uni-hamburg.de

[2]Leibniz-Institut für die Pädagogik der Naturwissenschaften und Mathematik, Kuhnkestr. 2, 24118 Kiel, Germany, ORCID: 0000-0002-6673-3528, mertens@leibniz-ipn.de

[3]Leibniz-Institut für die Pädagogik der Naturwissenschaften und Mathematik, Kuhnkestr. 2, 24118 Kiel, Germany, ORCID: 0000-0001-9714-6505, tjansen@leibniz-ipn.de

[4]Universität Hamburg, Sedanstraße 19, 20146 Hamburg, Germany, ORCID: 0000-0002-6564-9578, kira.weber@uni-hamburg.de

# AI, Expert or Peer?
# Provider Biases and Feedback Uptake Among Pre-Service Teachers

**Abstract**

The EU AI Act places teachers in charge of using high-risk AI safely in their classes, which requires them to assess AI-generated outputs. Feedback is one of the most consequential of these outputs, yet little is known about pre-service teachers' perceptions of AI-generated feedback. In a randomised experiment, 273 pre-service teachers each received one of 30 written feedback messages on a mathematics learning goal, produced under identical instructions by an expert, a peer, or a large language model (LLM). Without knowing the source, the participants judged who had written the message, rated six feedback perception subscales, and revised the learning goal. Source judgements were inaccurate (peer 46%, expert 40%, LLM 36%) and followed message length, not coded feedback quality. LLM feedback received more positive evaluations when ascribed to a human source. Ratings did not differ between feedback ascribed to experts and to peers. Relative uptake was highest for LLM feedback (52%). Coded feedback quality was the only variable significantly associated with uptake. Therefore, beliefs about who wrote a message shaped perceptions but not uptake. A feature-level analysis revealed that Valence was associated with perceptions only and an instructional composite with both perceptions and uptake. The implications for AI-related evaluative skills in teacher education are discussed.

## Practitioner Notes

### What is already known about this topic

- LLMs can generate feedback that meets or exceeds the quality of expert feedback.
- Nonetheless, students often underutilise LLM feedback, even when it is provided explicitly.
- Perceived source can influence perceptions of feedback, which may affect whether and how feedback is used.

### What this paper adds

- This paper reports on a randomised experiment with 273 pre-service teachers that examined how actual and perceived source and recognition accuracy shape perception and uptake.
- LLM feedback was perceived most positively when falsely ascribed to a human source.
- While perceived source was associated with feedback perceptions, only coded feedback quality was associated with uptake.
- Recognition accuracy of the feedback providers was generally low (peer ≈ 46%, expert ≈ 40%, and LLM ≈ 36%).

### Implications for practice and/or policy

- Feedback quality was associated with uptake and should guide the design and implementation of AI-generated feedback.
- Effectively integrating LLM feedback requires addressing learners' heuristics and biases.
- Hybrid feedback models that combine the scalability of LLMs with the credibility of humans may improve the perception and uptake of feedback.

# Introduction

As outlined in the AI Competency Framework for Teachers (UNESCO, 2024), teachers are expected to critically evaluate AI outputs and integrate them responsibly into instructional settings. One of the most relevant outputs is AI-generated feedback, as feedback has been shown to promote reflection, guide improvement, and enhance motivation (Hattie & Timperley, 2007; Narciss, 2013; Wisniewski et al., 2020). Yet feedback is only effective if students engage with it (Lipnevich & Smith, 2022; Panadero, 2023; Winstone et al., 2017). Engagement is influenced by students' perceptions of the feedback source due to social inference processes (i.e., inferring the provider's motives and expertise) and affective appraisal (i.e., evaluating feedback based on its personal relevance and emotional impact; Ilgen et al., 1979; Strijbos et al., 2021). In higher education, specifically teacher education, where pre-service teachers frequently receive feedback from experts, peers, and increasingly, AI, these perceptions are particularly important. This is not only relevant for their own learning but also as they prepare to use AI-driven feedback systems in their future teaching careers (Moorhouse & Kohnke, 2024; Prilop et al., 2024).

While the quality of AI feedback continues to improve, recent evidence demonstrates that, with appropriate prompting, AI feedback can match or even surpass expert feedback quality (Dai et al., 2024; Jacobsen & Weber, 2025). Nevertheless, some studies suggest that AI feedback, while having similar quality, is still perceived as less helpful or trustworthy compared to teacher feedback (Er et al., 2024; Jansen et al., 2024; Rubin et al., 2025; Ruwe & Kuklick, 2025; Steiss et al., 2024). Such negative perceptions may partially explain recent findings: Only half of university students use AI for feedback (Henderson et al., 2025), and only half of school pupils act upon the AI feedback they receive (Jansen et al., 2025). Notably, although students often hold distinct perceptions or biases about the quality and trustworthiness of feedback sources (Escalante et al., 2023; Nazaretsky et al., 2024; Ruwe & Mayweg-Paus, 2024), preventing effective feedback implementation (Lipnevich & Smith, 2022), empirical evidence indicates that pre- and in-service teachers frequently struggle to reliably distinguish between AI-generated and human-written texts (Fleckenstein et al., 2024).

Taken together, AI feedback is often underutilised, possibly due to negative perceptions of AI as a feedback source. This raises an important question: Do perception biases prevent the practical use of high-quality AI feedback? Addressing this issue requires research that examines not only whether teachers can identify feedback sources, but also how their source judgements, accurate or not, relate to feedback perception and to feedback uptake, that is, to the implementation of feedback suggestions in a revised product (Carless & Boud, 2018).

However, there is a need for research investigating both (a) whether teachers can correctly distinguish between large language model- (LLM), expert-, and peer feedback when sources are concealed, and (b) how the perceived feedback source relates to teachers' perceptions and subsequent uptake of the feedback. This study aims to address these questions and close this research gap with experimental data.

## Theoretical Background

### The Role of Feedback in Learning

Feedback can be understood as information that helps students identify and reduce the gap between their current level of knowledge or performance and the desired outcome of a learning process (e.g., Kluger & DeNisi, 1996; Narciss, 2008; Smith & Lipnevich, 2018). Contemporary approaches increasingly conceptualise feedback as a dialogical rather than purely transmissive process (Boud & Molloy, 2013; Carless, 2015; Griffiths et al., 2023; Lipnevich & Smith, 2022). In this process, meaning is actively constructed by students through social interaction and emotional appraisal (Ilgen et al., 1979; Strijbos et al., 2021). Henderson et al. (2025) describe this multidimensionality as including the informational content, the interactive processes involved, and the resulting cognitive and behavioural impacts on students. Therefore, understanding not only the nature of feedback itself, but also how students perceive, interpret, and respond to it, has become central to contemporary feedback research (Henderson et al., 2025). Engagement with feedback comprises cognitive, affective, and behavioural dimensions (Fredricks et al., 2004). The present study addresses the cognitive and affective dimensions through feedback perceptions and the behavioural dimension through uptake, which is operationalised as the implementation of suggestions in a revision task. The following section explores the key determinants that shape these processes systematically.

### Determinants of Feedback Effectiveness

The effectiveness of feedback depends on multiple interrelated factors, such as context, source, message, learner characteristics, processing, and outcomes, as described in the Student-Feedback Interaction Model (SFIM; Lipnevich & Smith, 2022). The focal point of the SFIM is the question of how students utilise feedback and the role of various factors in this process. This topic has become increasingly interesting to researchers in recent years, as evidenced by multiple systematic reviews and conceptual frameworks (Lipnevich & Panadero, 2021; Lipnevich & Van der Kleij, 2021; Panadero & Lipnevich, 2022; Winstone et al., 2017). According to the SFIM, students process feedback sequentially: they first evaluate it cognitively (i.e., perceived *fairness* and *usefulness*), respond affectively (i.e., *positive* or *negative affect*), and then form motivational intentions (i.e., *willingness to improve*), which ultimately translate into actions such as feedback uptake (Lipnevich & Smith, 2022; Strijbos et al., 2021).

Research in cognitive psychology has revealed that people often rely on mental shortcuts, known as heuristics, to simplify judgement and decisions (Gigerenzer & Gaissmaier, 2011). Heuristics are efficient strategies that reduce cognitive effort, especially in complex or uncertain situations. However, this efficiency comes at the cost of increased error proneness, as heuristics may lead to systematic deviations from normative judgement, known as biases. A prominent example is the expertise heuristic, whereby perceived expertise of the source disproportionately shapes perceptions (Go et al., 2014; Meinert & Krämer, 2022; Hovland et al., 1953).

These biases stem from what has been termed System 1 thinking, which is fast, intuitive, and automatic (Kahneman, 2011). While System 1 produces plausible reasons for actions, these are not necessarily the true causes of behaviour (Joughin et al., 2017; Kahneman, 2011; Stanovich & West, 2000). In contrast, System 2 operates in a deliberate, effortful, and orderly manner, supporting more analytical and reflective judgements (Kahneman, 2011). Biases ingrained in System 1 thinking may limit the impact of even high-quality feedback, which nevertheless is crucial for meaningful performance improvement (Ericsson et al., 1993).

Social identity theory offers a complementary explanation of source effects: when human providers are treated as an ingroup and AI as an outgroup, the operative distinction is human versus machine (Mertens et al., 2025; Tajfel & Turner, 1979).

In teacher education, acceptance of feedback and improvement in performance are closely tied to its quality (Prilop et al., 2021). However, even high-quality feedback may be disregarded if the source is perceived as untrustworthy or unqualified, as students form credibility judgements based on their perceptions of the feedback provider (Escalante et al., 2023; Zhang et al., 2025). Additionally, misidentifying the feedback provider can distort students' perceptions of its quality and utility, consequently affecting their willingness to act on it (Nazaretsky et al., 2024; Zhang et al., 2025). This suggests that identifying the feedback provider, whether correctly or incorrectly, is closely associated with how feedback is received and processed.

Thus, it is important to understand how students perceive and evaluate feedback from different providers, not only because (mis-)identifications of the source can bias the perceived quality (Nazaretsky et al., 2024), but also because the same feedback message can prompt different learning outcomes depending on the perceived source (Kao & Reynolds, 2024).

### Feedback Providers in Teacher Education

In teacher education, feedback is typically provided by experts, such as experienced educators, or by peers, such as fellow pre-service teachers (Lu, 2010; Kraft et al., 2018). Although expert feedback tends to be of higher quality (Prins et al., 2006; Prilop et al., 2019), its provision is limited by resource and time constraints (Demszky et al., 2023). Furthermore, hierarchical structures may discourage students from seeking further explanations (Carless, 2006; Winstone et al., 2017). Peer feedback, by contrast, is more accessible and promotes collaboration, though it frequently lacks the depth and specificity needed for substantial improvement (Prilop et al., 2020).

Intelligent tutoring systems (ITS) advanced individualised feedback but remain limited in depth, metacognitive support, and scalability (Aleven et al., 2016; Kulik & Fletcher, 2016). In contrast, LLMs enable more flexible and scalable feedback generation across diverse tasks and domains (Chiu et al., 2023; Steiss et al., 2024; Yin et al., 2024).

### Source Perception Effects

Research consistently shows that evaluations of the feedback source shape feedback perceptions, though the direction of these effects varies across studies. In general, users' mental models influence their perceptions of its trustworthiness, empathy, and effectiveness (Pataranutaporn et al., 2023). Regarding feedback specifically, some studies indicate higher trust towards AI feedback compared to teacher feedback (Ruwe & Mayweg-Paus, 2024), while others found that students' preferences between AI and human feedback were evenly split, with AI appreciated for clarity and precision, while human tutors were valued for interactivity and personalisation (Escalante et al., 2023). Similarly, Henderson et al. (2025) confirm that while higher education students appreciate AI feedback for its accessibility and clarity, they consistently ascribe greater expertise and authority to teacher feedback. Zhang et al. (2025) found that AI feedback was rated less favourably when participants were informed of its AI origin, whereas hybrid feedback (co-produced by AI and humans) received the highest ratings. Nazaretsky et al. (2024) demonstrated that feedback misidentified as human was perceived more positively than feedback correctly ascribed to AI, revealing a consistent pro-human bias in perceptions. Mertens et al. (2025) tested this asymmetry by attaching randomly assigned source labels to AI-produced messages. Teachers rated the AI-labelled version as less credible, useful, and fair. 73% preferred the teacher-labelled message. However, source misidentification is not universally detrimental to AI feedback. Kao and Reynolds (2024) found that students who believed their feedback originated from an AWE system performed better on second language writing tasks than those who assumed it came from a teacher. High-quality feedback, regardless of its actual source, tends to be ascribed to expert human sources, indicating systematic biases that conflate perceived quality with assumptions about the source (Escalante et al., 2023). This pattern suggests that students use quality cues to infer source identity, creating a circular relationship between perceived source and perceived quality. Recognition accuracy (i.e., the degree to which students correctly identify the true feedback provider) thus emerges as a potentially critical but understudied factor in feedback effectiveness (Nazaretsky et al., 2024). Despite this emerging evidence, important questions remain about whether and how biases against automated feedback actually influence the way students process and enact feedback (Henderson et al., 2025; Nazaretsky et al., 2024). It is unclear whether biases merely alter perceptions, or also impact how students use and benefit from feedback in practice (Henderson et al., 2025).

This review reveals important gaps in understanding how source perception relates to the use of feedback in teacher education. Most prior studies disclosed or directly manipulated the provider label. The beliefs that students form on their own about the source, and their role in feedback processing, remain underexplored. The relationship between source (mis-)identification, feedback perceptions, and feedback uptake (i.e., implementation of feedback suggestions; Carless & Boud, 2018) remains unclear.

## Aim of the Research and Research Questions

Although LLMs can generate feedback that is comparable to or better than human feedback (Dai et al., 2024; Jacobsen & Weber, 2025) only half of the students use LLM feedback (Henderson et al., 2025) even after receiving it explicitly (Jansen et al., 2025). Research shows that perceptions of the feedback source, especially whether students believe it stems from a human or a LLM, influence how the feedback is trusted and evaluated (Escalante et al., 2023; Nazaretsky et al., 2024; Ruwe & Kuklick, 2025; Ruwe & Mayweg-Paus, 2024; Zhang et al., 2025). Subsequently, these perceptions may influence how much the feedback is utilised (Ilgen et al., 1979; Lipnevich & Smith, 2022; Panadero, 2023; Strijbos et al., 2021; Winstone et al., 2017).

Despite these insights, little is known about how the perceived source relates to feedback perceptions and uptake. This could explain the low observed usage of LLM feedback. This study addresses this gap by examining how pre-service teachers perceive and act upon feedback that was written by a LLM, an expert or a peer. By focusing on pre-service teachers' perceptions, this experimental study offers new insights into how source (mis-)identifications can shape cognitive, emotional, and behavioural responses to feedback. Drawing on the SFIM and human–AI interaction research, the study focuses on the relations of perceived source, recognition accuracy, and coded feedback quality with cognitive, affective, and behavioural responses to feedback.

Designs that attach source labels to identical messages isolate the effect of source information on the evaluation of equivalent content (Mertens et al., 2025; Nazaretsky et al., 2024; Ruwe & Kuklick, 2025). Message-feature approaches, in turn, compare disclosed sources and decompose their effects into properties of the messages (Weidlich et al., 2026). The present design builds on these approaches and addresses an additional question: when source information is withheld, which sources do pre-service teachers infer from the message itself, and do these inferences or the message characteristics influence perception and uptake? This question gains relevance because AI-generated text is not reliably identifiable (Fleckenstein et al., 2024) and feedback can be co-produced with AI (Zhang et al., 2025), so that source labels cannot be verified and beliefs about the source formed from message characteristics can play a key role in response to feedback.

Hence, the following research questions are addressed:

**Research Question 1** How are pre-service teachers' perception and uptake of feedback messages associated with the perceived source of the feedback (LLM vs. human [expert vs. peer])?

**Hypothesis 1** Based on prior research on feedback source perceptions (e.g., Nazaretsky et al., 2024; Zhang et al., 2025) and expertise heuristics (Go et al., 2014; Meinert & Krämer, 2022; Hovland et al., 1953), we hypothesise that feedback perceived as coming from an expert will be rated more favourably across cognitive and affective perception dimensions than feedback perceived as coming from an LLM or a peer.

**Hypothesis 2** Furthermore and drawing on the SFIM (Lipnevich & Smith, 2022), we expect that perceived source will also be related to feedback uptake, with feedback ascribed to experts being more likely to be integrated into pre-service teachers' revised learning goals.

**Research Question 2** How is this association moderated by pre-service teachers' accuracy in recognising the actual source, while accounting for coded feedback quality?

**Hypothesis 3** We formulate the hypothesis that correctly recognising the actual source will amplify the effects of source on perception and uptake. This is derived from prior evidence showing links between recognition accuracy, AI literacy, and feedback perceptions (e.g., Ruwe & Mayweg-Paus, 2024; Zhang et al., 2025).

**Hypothesis 4** Moreover and in line with feedback quality models (Gielen et al., 2010; Narciss, 2013), we argue that coded feedback quality will independently predict both perception and uptake, beyond perceived source and recognition accuracy.

# Methods

The study complied with APA ethical standards and EU data protection regulations. Participation was voluntary, informed consent was obtained, and the university's data protection officer approved the procedures. As none of the conditions requiring ethics approval under DFG guidelines applied, no formal approval was necessary.

## Sample and Procedure

The sample consisted of 273 bachelor's students in their fourth semester of a teacher education programme at a German university. All participants were preparing to become primary or secondary school teachers and completed a four-week teaching practicum at the end of that semester. Age and gender information was available for 242 students ($M_{\text{age}}$ = 22.65, $SD_{\text{age}}$ = 3.29; 81.0% female); for 31 students, this demographic data was not provided. Most students reported having no (37.6%) or only limited prior teaching experience (34.7% with 1–10 hours; 12.0% with 11–30 hours; 14.1% with more than 30 hours of independent teaching). Data from two cohorts, those enrolled in the summer semesters of 2023 and 2024, were combined for analysis.

To ensure ecological validity, the study context closely mirrored the experience of pre-service teachers during their practicum. In this phase, all participants were required to independently teach their first lessons and submit detailed lesson plans, including explicit learning goals. Prior to conducting one of their lessons, students received written feedback, predominantly provided by student assistants owing to the educators' workload, on their learning goals and lesson plans. Typically, each student received feedback only once during the practicum, despite preparing multiple lessons, due to time constraints from educators in the module. The study procedure thus replicated an instructionally consequential and infrequent feedback scenario characteristic of teacher education, with participants engaging with written feedback on a hypothetical mathematics learning goal.

Participants were first presented with the following learning goal: *Students should be able to recognise a right-angled triangle and understand the Pythagorean theorem.* The learning goal incorporated three mistakes (i.e., no activity verb; instructional rather than learning goal; and multiple learning goals in a single statement). Each participant was randomly assigned one feedback message (feedback regarding the above stated learning goal) drawn from a pool of 30 pieces of feedback with 10 expert-generated feedback messages, 10 peer-generated feedback messages and 10 LLM-generated feedback messages.

Expert feedback was written by educational scientists ($n$ = 3), professors of education ($n$ = 2), teachers ($n$ = 4) and one teacher educator, each contributing one message. Peer feedback was written by 10 students in the final phase of a bachelor's teacher education programme, corresponding to the participant population. Participation was voluntary. LLM feedback was generated with ChatGPT-4, based on a prompt tested in prior research on LLM feedback (Jacobsen & Weber, 2025; Jacobsen et al., 2026). All three provider groups received the identical written prompt, including the criteria for learning goals, the specifications regarding tone and first-person phrasing, and the constraint of 150 to 250 words (see Figure S1 in the supplementary material for the full prompt). No provider-specific adaptations were made, and no selection among messages took place. Despite this, the resulting messages differed in length (LLM: $M$ = 233.6 words, $SD$ = 74.4; expert: $M$ = 248.9, $SD$ = 84.5; peer: $M$ = 129.9, $SD$ = 93.6). This procedure was chosen to capture the feedback produced by each provider group under identical instructions, rather than to standardise length, structure, or tone across sources. Differences in message characteristics therefore reflect what each provider group produced under equivalent instructions.

After reviewing the assigned feedback (see Figure S2 in the supplementary material for examples of feedback messages), participants were required to identify its source by selecting one of the following options: LLM, expert, or peer. Subsequently, participants were asked to complete the Feedback Perception Questionnaire (FPQ; Strijbos et al., 2021). The pre-service teachers were asked to rate their perceptions as if they had received the feedback themselves. In the final step, participants were asked to reformulate the given learning goal based on the feedback they had received. To ensure alignment with the feedback provided, both the original learning goal and the assigned feedback were displayed on the screen as participants formulated their revised learning goal. The task was: "Please reformulate the learning goal based on the feedback you received. Note that the task is not to implement your own idea of a learning goal, but to implement the suggestions from the feedback."

## Instruments and measures

### Feedback perceptions

Participants evaluated their perceptions using the full 18-item version of the FPQ (Strijbos et al., 2021). The questionnaire was administered in German translation and assessed various dimensions of feedback perception. Cognitive perceptions included *fairness* (e.g., “I consider this feedback fair”; α = .87), *usefulness* (e.g., “I consider this feedback useful”; α = .93), and *acceptance* (e.g., “I accept this feedback”; α = .86). Motivational perceptions encompassed *willingness to improve* (e.g., “I am willing to improve my performance because of the feedback”; α = .84), as well as *positive* (e.g., “I would feel confident if I received this feedback on my revision”; α = .90) and *negative affect* (e.g., “I would feel frustrated if I received this feedback on my revision”; α = .88). Participants were instructed as follows: “Imagine you have just received the feedback you have just read. Please rate the following statements in relation to the feedback.”. They rated each item using a slider scale ranging from 1 (does not apply at all) to 10 (absolutely applies). This scenario-based administration corresponds to the validation of the FPQ, in which participants rated feedback presented within a hypothetical scenario (Strijbos et al., 2021). Scenario-based ratings have been shown to evoke reactions comparable to real situations (Robinson & Clore, 2001). Negatively phrased items were reverse-coded to ensure consistency in data interpretation. All subscales were operationalised as the mean of relevant items. Based on correlations identified by Strijbos et al. (2021), the subscales were grouped into a cognitive dimension (fairness, usefulness, acceptance) and a motivational dimension (willingness to improve, affect). The perception models are reported at the subscale level to ensure that the pattern across dimensions remains visible. The composites enter the uptake model, where they represent participants’ perceptions as predictors of implementation.

### Feedback uptake

In this study, the term uptake refers to the implementation of feedback suggestions in the revised learning goal. Feedback uptake was operationalised using two metrics. At the participant level, the correction rate represented the proportion of suggested improvements implemented by each participant and served as the primary outcome in inferential analyses. At the group level, an aggregated uptake index was calculated by dividing all implemented suggestions by the total number of recognised errors per provider. This allowed for standardised comparison of uptake across LLM, expert, and peer feedback conditions.

### Feedback quality

Coded feedback quality, served as a covariate in all inferential models to control for objective differences when modelling the effects of source (mis-)identification and perception on feedback uptake. To examine the quality of the feedback, a quantitative content analysis was conducted. The coding was based on a theory-informed manual by Prilop et al. (2019) that builds on the model by Prins et al. (2006). It has since been adapted to LLM feedback and has been used in two previous studies (Jacobsen & Weber, 2025; Jacobsen et al., 2026). The current version comprises nine categories: *Assessment criteria*, *Specificity*, *Explanation*, *Presence of suggestions for improvement*, *Explanation of suggestions*, *Errors*, *Questions*, *First person* and *Valence*. Each feedback message was treated as a unit of analysis and every category was rated on a three-point scale (0 = low quality, 1 = moderate quality, 2 = high quality). Two trained raters independently coded one third of the messages ($n$ = 10; κ = .63–1.00); the remaining messages ($n$ = 20) were coded by one rater. For detailed definitions of the coded feedback quality categories, kappa values and examples of good feedback for each category, see supplementary Table S1.

LLM feedback was rated highest in *Assessment criteria*, *Presence of suggestions for improvement*, *First person*, and *Valence* (see Figure 1). Expert feedback was rated higher than the other providers in giving *Explanations* and posing *Questions*, while peer feedback consistently received the lowest ratings across most categories.

**Figure 1.** *Mean Quality Ratings for each Coded Feedback Quality Category by Provider (LLM, Expert, Peer), with Error Bars representing Standard Deviations*

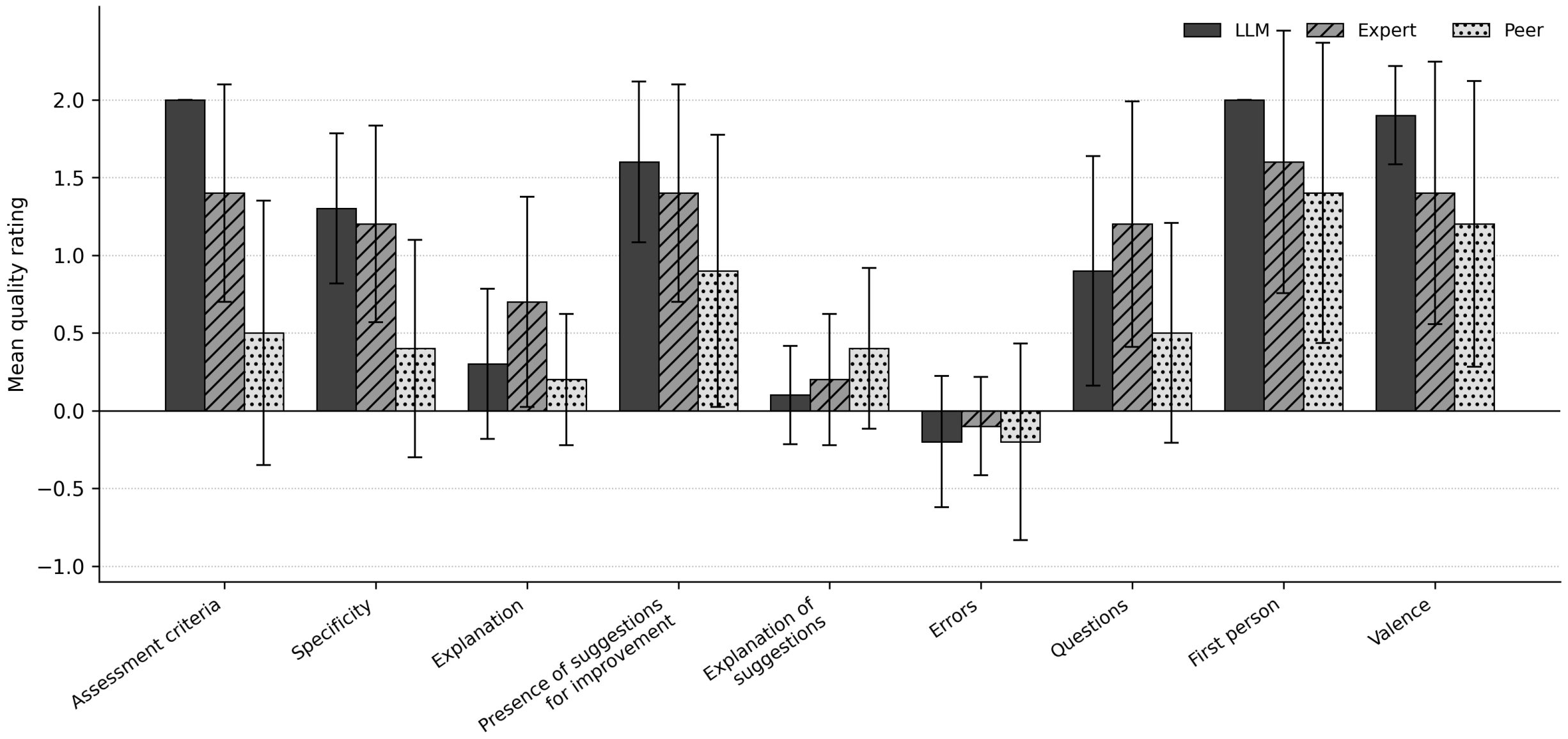

*Note. Bars show means across the ten messages per provider; error bars show standard deviations. Assessment criteria and First person carry no error bar for LLM feedback because all ten messages received the maximum code. The Errors category is reverse-scored (0 = no content errors, −1 = one content error, −2 = several content errors).*

### Recognition accuracy

Recognition accuracy (i.e., whether participants correctly identified the true source of each feedback message) was operationalised as a binary variable indicating whether the perceived source matched the actual source (1 = correct, 0 = incorrect). For each participant and condition (LLM, expert, peer), mean accuracy scores were calculated. Descriptive statistics summarised participants' ability to identify the source of the feedback correctly. The resulting accuracy rates and the pattern of misidentifications are reported in the Results section.

## Data Analyses

### Perceived Source and Feedback Quality

To investigate whether the perceived source of feedback was associated with its quality, a one-way ANOVA was conducted with coded feedback quality as the dependent variable and the perceived source (1 = LLM, 2 = expert, 3 = peer) as the between-subjects factor.

### Feedback Perception

All analyses were conducted using R (version 4.5.0; R Core Team, 2023). Linear and binomial regression models examined how the perceived source of feedback (LLM, expert, peer) and pre-service teachers' recognition accuracy were associated with their perception and uptake of feedback. All models included coded feedback quality as a covariate. The peer feedback condition was used as the reference category. Each participant was randomly assigned one of the 30 feedback messages, so that the actual source varied randomly across participants. Participants reported the perceived source, and recognition accuracy was derived by comparing this with the actual source. Therefore, analyses involving the perceived source or recognition accuracy are correlational. All models were estimated on complete cases. Of the 273 participants, 262 provided a source judgement, of whom 231 reported their age and gender. Together with missing item-level values on the outcome scales, this yields $N = 222$ to 227 depending on the outcome. The uptake model also required the number of recognised and implemented suggestions and is therefore based on data from 160 participants.

### Effect of feedback provider on feedback perceptions

A series of multiple linear regression analyses was conducted to examine how the perceived source of feedback and participants' recognition accuracy were associated with their responses to feedback. Each model included the actual source, recognition accuracy (correct vs. incorrect identification), and their interaction, i.e., whether the effect of the actual source differs depending on recognition accuracy, while controlling for age, gender, and coded feedback quality.

### Feedback Uptake

Prior to analysis, 13 cases with missing or invalid values (e.g., unchanged learning goal) were excluded. The analysis proceeded in three steps: First, absolute and relative frequencies of errors recognised by each feedback provider (LLM, expert, peer) were calculated. Second, the number and proportion of suggested corrections that pre-service teachers implemented, disaggregated by provider, was assessed. Third, a relative uptake index (i.e., correction rate) was computed by dividing the total number of implemented suggestions by the total number of recognised errors for each provider. The correction rate accounted for variations in recognition rates and provider group sizes, yielding a standardised metric for comparing relative feedback uptake across conditions.

Because the correction rate is a proportion bounded between 0 and 1, feedback uptake was modelled using binomial logistic regression. For each participant, the number of implemented suggestions relative to the number of suggestions identified in the assigned feedback message served as the outcome. This specification accounts for the bounded outcome and weights participants by the number of suggestions received.

## Results

### Feedback Perceptions

Descriptive analyses addressing the first research question on how the perceived source of feedback (LLM versus human, either expert or peer) relates to pre-service teachers' perception and uptake of feedback messages revealed a systematic pattern in feedback perceptions. Misidentifying feedback sources was associated with differences in these evaluations, with higher ratings when feedback was ascribed to human sources and lower ratings when it was ascribed to the LLM (Table 1).

**Table 1.**

*Means and Standard Deviations for all Subscales of the Feedback Perceptions of Actual and Perceived Source*

| | **LLM Feedback** | | | | **Expert Feedback** | | | | **Peer Feedback** | | | |
|---|---|---|---|---|---|---|---|---|---|---|---|---|
| | Actual | | Perceived | | Actual | | Perceived | | Actual | | Perceived | |
| | *n* = 85 | | *n* = 76 | | *n* = 95 | | *n* = 81 | | *n* = 82 | | *n* = 105 | |
| | *M* | *SD* | *M* | *SD* | *M* | *SD* | *M* | *SD* | *M* | *SD* | *M* | *SD* |
| Fairness (3 items) | 7.90 | 1.64 | 6.52 | 2.02 | 7.42 | 1.82 | 7.95 | 1.48 | 6.61 | 1.92 | 7.43 | 1.81 |
| Usefulness (3 items) | 7.52 | 2.10 | 5.87 | 2.26 | 7.24 | 1.95 | 7.91 | 1.79 | 5.98 | 2.27 | 7.00 | 2.08 |
| Feedback Acceptance (3 items) | 8.25 | 1.89 | 6.90 | 2.19 | 7.95 | 1.82 | 8.48 | 1.47 | 6.96 | 2.04 | 7.78 | 1.95 |
| Willingness to Improve (3 items) | 7.64 | 1.89 | 6.82 | 1.89 | 7.64 | 1.61 | 7.93 | 1.57 | 6.84 | 2.01 | 7.38 | 1.96 |
| Positive Affect (3 items) | 5.69 | 1.95 | 4.74 | 2.10 | 5.47 | 1.92 | 5.79 | 1.96 | 4.82 | 2.12 | 5.42 | 1.91 |
| Negative Affect (3 items) | 1.93 | 1.44 | 2.73 | 1.69 | 2.63 | 1.91 | 2.32 | 1.84 | 2.47 | 1.69 | 2.10 | 1.60 |

**Perceptions by actual source and recognition accuracy**

The first set of models covered the entire sample (*N* = 222 to 227 depending on the outcome; see Table 2), regressing each perception scale on the actual source (expert, peer or LLM), whether participants correctly identified that provider, the interaction between the two and coded feedback quality, age and gender. Peer feedback and correct identification served as reference categories. For fairness, a significant interaction between LLM feedback and recognition accuracy ($b$ = 1.29, $SE$ = 0.59, $t$ = 2.20, $p$ = .029) indicated that fairness ratings were higher when LLM-generated feedback was perceived as being provided by a human rather than by an LLM. The same pattern emerged for usefulness ($b$ = 1.82, $SE$ = 0.67, $t$ = 2.73, $p$ = .007) and for acceptance ($b$ = 1.26, $SE$ = 0.61, $t$ = 2.06, $p$ = .041). For willingness to improve the coefficient was of comparable size and did not reach significance ($b$ = 1.05, $SE$ = 0.59, $t$ = 1.78, $p$ = .077; see Table 2). Coded feedback quality was a significant covariate in the models for fairness, usefulness and acceptance (all $p$s $\leq$ .012); the interaction estimates reported above are therefore adjusted for quality. Within the same models, among participants who identified the provider correctly, expert-generated feedback received higher ratings than peer-generated feedback for usefulness ($b$ = 1.26, $SE$ = 0.53, $t$ = 2.38, $p$ = .018) and acceptance ($b$ = 1.18, $SE$ = 0.48, $t$ = 2.43, $p$ = .016).

Additionally, linear regression models were used to examine whether emotional responses to feedback were associated with the actual source and recognition accuracy, with their interaction as a predictor. Neither feedback source, recognition accuracy, nor their interaction was significantly associated with positive affect (all $p$s $\geq$ .083) or negative affect (all $p$s $\geq$ .058). Coded feedback quality was associated with both (positive affect: $b$ = 1.29, $p$ = .009; negative affect: $b$ = −1.09, $p$ = .007).

**Perceptions by perceived source within the LLM condition**

Recognition accuracy is binary and pools participants who ascribed LLM-generated feedback to an expert with those who ascribed it to a peer. A second set of models separated these two groups. These models were restricted to the 85 participants who received LLM-generated feedback. The binary accuracy indicator was replaced with a three-level predictor of the perceived source (perceived expert, $n$ = 23; perceived LLM, $n$ = 31; perceived peer, $n$ = 31), with correct identification as the reference category. Coded feedback quality was included as a covariate and missing values at item level left 83 to 84 cases per model. In this second set, the predictor refers to participants' judgements, whereas in the first set, the provider coefficients refer to the actual, assigned provider.

Ratings were higher when the feedback was perceived to come from a peer than when it was identified as LLM-generated (usefulness: $b$ = 1.22, $SE$ = 0.49, $p$ = .015; willingness to improve: $b$ = 1.03, $SE$ = 0.47, $p$ = .033). The coefficient for perceived expert was comparable in size (usefulness: $b$ = 1.00, $SE$ = 0.57, $p$ = .085), and feedback perceived to come from experts and from peers did not differ (all $p$s $\geq$ .57; see supplementary Table S2). Across both sets of models, the difference was between the human and the LLM categories. Sensitivity analyses controlling for message length reproduced the interaction effects of the first set and the perceived-source contrasts of the second (see supplementary Table S3).

**Table 2.**

*Linear Model Results for Perceptions of Fairness, Usefulness, Acceptance and the Willingness to Improve*

| | **Fairness** (*N* = 227) | | | | **Usefulness** (*N* = 227) | | | | **Acceptance** (*N* = 225) | | | | **Willingness to Improve** (*N* = 222) | | | |
|---|---|---|---|---|---|---|---|---|---|---|---|---|---|---|---|---|
| | *b* | *SE* | *t* | *p* | *b* | *SE* | *t* | *p* | *b* | *SE* | *t* | *p* | *b* | *SE* | *t* | *p* |
| Intercept | **6.155** | **0.466** | **13.217** | **< .001** | **5.178** | **0.523** | **9.897** | **< .001** | **6.552** | **0.478** | **13.695** | **< .001** | **6.447** | **0.471** | **13.696** | **< .001** |
| LLM | -0.012 | 0.514 | -0.023 | .981 | -0.160 | 0.583 | -0.275 | .784 | -0.098 | 0.537 | -0.182 | .856 | -0.140 | 0.516 | -0.272 | .786 |
| Expert | 0.812 | 0.467 | 1.738 | .084 | **1.264** | **0.530** | **2.384** | **.018** | **1.177** | **0.484** | **2.430** | **.016** | 0.735 | 0.470 | 1.563 | .120 |
| Incorrect guess | -0.243 | 0.408 | -0.595 | .552 | -0.270 | 0.463 | -0.582 | .561 | -0.041 | 0.423 | -0.098 | .922 | 0.152 | 0.413 | 0.367 | .714 |
| Coded feedback quality | **1.086** | **0.426** | **2.551** | **.011** | **1.277** | **0.481** | **2.656** | **.008** | **1.112** | **0.440** | **2.528** | **.012** | 0.697 | 0.432 | 1.616 | .108 |
| Age (scaled) | -0.101 | 0.114 | -0.891 | .374 | -0.184 | 0.129 | -1.428 | .155 | 0.016 | 0.118 | 0.135 | .893 | -0.197 | 0.114 | -1.723 | .086 |
| Gender (male) | 0.027 | 0.309 | 0.087 | .931 | -0.308 | 0.347 | -0.888 | .375 | 0.178 | 0.317 | 0.560 | .576 | 0.032 | 0.307 | 0.104 | .917 |
| LLM x incorrect guess | **1.288** | **0.586** | **2.200** | **.029** | **1.819** | **0.665** | **2.733** | **.007** | **1.258** | **0.611** | **2.057** | **.041** | 1.051 | 0.592 | 1.776 | .077 |
| Expert x incorrect guess | -0.735 | 0.566 | -1.299 | .195 | -0.942 | 0.644 | -1.464 | .145 | -0.925 | 0.589 | -1.569 | .118 | -0.497 | 0.575 | -0.864 | .389 |
| Adjusted $R^2$ | .136 | | | | .179 | | | | .135 | | | | .074 | | | |

*Note.* Intercept = Peer-generated feedback as reference. Adjusted $R^2$ is reported to account for the model's complexity. Significance indicated by bold formatting.

### Recognition Accuracy

Descriptive analyses related to the second research question examined how the relationship is moderated by pre-service teachers' accuracy in recognising the actual source, accounting for coded feedback quality. Recognition accuracy varied by actual source. Peer feedback was identified most accurately (46%), followed by expert feedback (40%). LLM feedback was detected only slightly above chance (36%). Most misidentifications involved expert feedback mistaken for peer ($n$ = 36) and LLM feedback misidentified as peer ($n$ = 31). In total, incorrect identifications ($n$ = 155) outnumbered correct ones ($n$ = 107). To explore potential reasons, mean coded feedback quality was compared across the three perceived sources. The difference was not significant, $F(2, 259) = 2.54$, $p = .081$, $\eta^2 = .019$, and none of the Bonferroni-adjusted pairwise comparisons reached significance (all $p$s $\geq .075$), giving little indication that participants inferred the source from the quality of the message. Message length, by contrast, was associated with source judgements: messages judged to be expert-written were longer than messages judged to be LLM-generated ($b$ = 38.69, $SE$ = 14.56, $p$ = .008; see supplementary Table S3). Within the LLM condition, where all messages came from the same pool, the same pattern held ($b$ = 36.90, $SE$ = 15.98, $p$ = .024).

### Feedback Uptake

Regarding Research Question 1, descriptive differences in the number of mistakes identified and the proportion of implemented suggestions across feedback sources were examined first. Feedback generated by the LLM identified the most mistakes and showed the highest absolute and relative uptake in participants, followed by expert feedback and, lastly, peer feedback. Since peer feedback typically flagged only one issue, its potential for uptake was limited. A detailed summary of recognised mistakes, implemented suggestions, and standardised uptake ratios is provided in Table 3. Bonferroni-adjusted post hoc comparisons for recognised mistakes and implemented suggestions are reported in supplementary Table S4.

**Table 3.**

*Feedback Uptake by Actual Source*

| Provider | Number of Feedbacks | Mistakes recognised | Suggestions for improvement implemented | Mistakes recognised per feedback | Suggestions for improvement implemented per feedback | Relative Feedback Uptake |
|---|---|---|---|---|---|---|
| LLM | 85 | 198 | 103 | 2.3 | 1.2 | 52% |
| Expert | 95 | 211 | 96 | 2.2 | 1.0 | 45% |
| Peer | 82 | 106 | 44 | 1.2 | 0.5 | 42% |

To examine Research Question 2, a binomial logistic regression model was estimated to test whether the relationship between the actual source and uptake was moderated by recognition accuracy, while accounting for coded feedback quality and participants' feedback perceptions. Aggregated variables were used to include perceptions in the regression model (Strijbos et al., 2021). The means and standard deviations for these aggregated variables are presented in supplementary material Table S5.

The results showed that coded feedback quality was the only significant predictor of uptake ($b$ = 1.84, $SE$ = 0.56, $z$ = 3.30, $p$ = .001), with higher-quality feedback being more likely to be implemented. In contrast, neither the actual source, recognition accuracy, nor their interaction significantly predicted uptake (all $p$s ≥ .090). Participants' aggregated cognitive and affective perceptions were also unrelated to uptake ($p$s = .88 and .86). The model accounted for a small share of the variation in implementation (Nagelkerke $R^2$ = .11). However, only 160 observations were included in this model.

A feature-level analysis entered the coded Valence category, as an indicator of tone, and an instructional composite alongside the provider (supplementary Table S6). The provider coefficients for fairness and usefulness dropped to non-significance (all $p$s ≥ .124), whereas Valence was associated with fairness ($b$ = 0.29, $SE$ = 0.13, $p$ = .024) and usefulness ($b$ = 0.31, $SE$ = 0.14, $p$ = .029), and the instructional composite was associated with fairness ($b$ = 0.33, $SE$ = 0.15, $p$ = .029), usefulness ($b$ = 0.54, $SE$ = 0.17, $p$ = .002) and, most strongly, with uptake ($b$ = 0.71, $SE$ = 0.19, $p$ < .001).

## Discussion

This study investigated how the (mis-)identification of feedback sources relates to pre-service teachers' perceptions and their uptake of feedback. Judgements about the source were closely associated with perceptions, whereas uptake was related primarily to the quality of the message. The judgements themselves tracked message length. Coded feedback quality was not significantly associated with them.

The results are partly consistent with the first hypothesis. Feedback perception was associated with the perceived source. LLM feedback was rated as fairer, more useful and more acceptable when it was ascribed to a human source. This supports the SFIM proposition that contextual cues bias early cognitive evaluation. The present data extend the model by showing that a human label attached to LLM-generated feedback is enough to raise those ratings. Ratings did not differ between feedback ascribed to experts and to peers. The contrast therefore runs between human and machine sources, which fits the social identity account outlined above (Tajfel & Turner, 1979). The pattern matches studies in which perception ratings for LLM feedback fell once the source was disclosed (Nazaretsky et al., 2024; Zhang et al., 2025). It also matches experimental evidence in which an LLM label lowered credibility, usefulness and fairness ratings of an identical message (Mertens et al., 2025). Those designs assign a label. Here the source was inferred from unlabelled messages that the providers had written themselves. The present data allow the same comparison within the LLM-generated messages alone. There every message comes from the same pool and coded feedback quality is accounted for. Differences between the providers' messages therefore cannot explain the effect.

Coded feedback quality did not differ significantly across the three perceived sources. Because the perceived source was reported and not manipulated, an association in either direction is possible. Message characteristics indicative of quality may equally inform source judgements (Escalante et al., 2023). The design does not separate the two directions. This reservation concerns the comparison between the provider groups, whose messages differ. It does not concern the comparison within the LLM-generated messages, where a single set of messages was evaluated. Large-scale survey data also suggest that students often distrust LLM feedback, citing concerns about its reliability and authenticity (Henderson et al., 2025). Judgements of trust and judgements of quality are difficult to separate in students' evaluations of feedback (Zhang et al., 2025).

Contrary to the second hypothesis, LLM feedback achieved the highest relative uptake (52%), ahead of expert feedback (45%) and peer feedback (42%). This comparison refers to the provider who wrote the message, not to the source participants perceived. In the uptake model the perceived source was unrelated to uptake. The source therefore mattered for the immediate evaluation of feedback (Dijks et al., 2018; Ilgen et al., 1979; Kluger & DeNisi, 1996; Lipnevich & Smith, 2022; Strijbos et al., 2010). Uptake, by contrast, followed the quality of the message (Gielen et al., 2010; Narciss, 2013).

Contrary to the third hypothesis, correct recognition did not amplify the source effects. For perceptions the interaction ran in the opposite direction. Ratings of LLM-generated feedback were higher when the source was not recognised. For uptake, once coded feedback quality was accounted for, neither recognition accuracy, the perceived source, nor cognitive and affective perceptions was associated with implementation. Dual-process theory (Kahneman & Tversky, 1979; Tversky & Kahneman, 1989) helps to explain this distinction. The perceived source operates as a framing cue in the editing phase, shaping early, heuristic judgements. In contrast, the evaluation phase involves an analytic process, in which coded feedback quality is the variable most strongly associated with uptake. The dissociation corresponds to the discrepancy between perception and revision described by Weidlich et al. (2026). There perceptions followed the tone of the message, while the quality of the revision followed targeted, instructionally rich guidance. In this study, beliefs about who wrote a message were related to how it was judged but not to what participants did with it.

A feature-level analysis locates this dissociation in the properties of the messages. Entering the coded Valence category and the instructional composite alongside the provider reduced the provider coefficients for fairness and usefulness to non-significance. Valence was associated with perceptions only. The instructional composite was associated with both perceptions and uptake, and most strongly with uptake (supplementary Table S6). What distinguishes the providers in these data is therefore what their messages contain, and the two kinds of content act on different outcomes.

In support of the fourth hypothesis, coded feedback quality was associated with both outcomes. It was a significant covariate in the perception models for fairness, usefulness and acceptance, and the only significant predictor of uptake. This relates to the Message component of the SFIM framework (Lipnevich & Smith, 2022) and underlines the role of feedback quality in uptake (Gielen et al., 2010; Narciss, 2013). Despite the variation in perceptions that accompanied misidentifications, pre-service teachers implemented feedback in line with its quality.

Pre-service teachers had difficulty identifying feedback sources. Peer feedback was correctly identified most frequently (46%), followed by expert (40%) and LLM feedback (36%). This aligns with findings that students struggle to distinguish AI- from human-generated text (Fleckenstein et al., 2024). Students often rely on unreliable cues such as linguistic warmth (Jakesch et al., 2023) or, in these data, message length.

Integrating LLM feedback into teacher education therefore requires addressing learners' heuristics and biases. This underlines the need for AI-related evaluative skills, as outlined in the UNESCO AI Competency Framework (2024). Information alone may not suffice. Explaining how an LLM produces its output did not change teachers' evaluations of LLM-labelled feedback (Mertens et al., 2025). Programmes could instead train pre-service teachers to evaluate feedback against explicit quality criteria such as those used here. Learners could compare AI-generated and human feedback on the same artefact, with the sources disclosed only after an initial evaluation. This makes source-related biases observable. Structured reflection on the reasons for trusting or rejecting a source could connect these exercises to repeated exposure and iterative feedback cycles. Hybrid feedback models combining the scalability of LLMs with human credibility may offer a way forward.

## Limitations and Directions for Future Research

When interpreting the results of this study, several limitations should be noted. First, the design involved comparing feedback produced by each provider group under identical instructions. While this approach reflects the feedback that pre-service teachers encounter in practice, it implies that message characteristics, such as length and the number of suggestions, varied with the provider. Coded feedback quality was included as a covariate in all models, and a sensitivity analysis controlling for message length is reported in the supplementary material. In that analysis the interaction effects remained comparable in size and significance, and coded feedback quality remained the only predictor of uptake. A feature-level analysis further indicated that the differences between providers in perception are carried by the content of the messages rather than by the provider label (supplementary Table S6). Nevertheless, comparisons by actual source capture provider groups as bundles of source and message characteristics rather than the isolated causal effect of a source label. Second, because the perceived source was reported rather than manipulated, associations between the perceived source and evaluations are correlational and consistent with bidirectional processes, as message characteristics may have informed source judgements. Third, participants evaluated feedback on a learning goal they had not authored and rated their anticipated responses. This procedure follows the scenario-based administration of the FPQ (Strijbos et al., 2021), which offers control of contextual variables and thereby supports internal validity (Atzmüller & Steiner, 2010), and it held the evaluated performance constant across participants. However, it may attenuate affective responses, which could contribute to the absence of source effects on affect, and it positions uptake as implementation of suggestions on a provided artefact rather than as engagement with feedback on one's own work. Fourth, the instructional design restricted variance by limiting the learning goal to three errors, constraining the predictive power of the uptake model, which was further based on a reduced sample ($n$ = 160). Fifth, despite receiving identical prompts, coded feedback quality varied across providers, especially for peer feedback, which was often shorter and contained fewer suggestions for improvement. The conditions were therefore not fully comparable in quality, and the results concerning source should be interpreted in conjunction with the quality covariate. Sixth, the study did not assess individual differences such as feedback or AI literacy, despite growing evidence that these factors relate to students' feedback responses (Hawkins et al., 2025; Henderson et al., 2025). Finally, the study focused on one-way written feedback in response to a single task.

Building on the findings of this study, future studies could examine feedback on self-authored work, extended feedback interactions, or replicate the findings with mental-model manipulations, such as framing AI motives, to test the robustness of heuristics and AI-related biases (Pataranutaporn et al., 2023). Longitudinal and repeated-measures designs could track changes over time, and multimodal feedback formats (e.g., audio, video) may offer additional insight into perception and engagement.

## Conclusion

This study provides empirical evidence of a bias favouring human sources over LLMs among pre-service teachers. LLM feedback received higher perception ratings when ascribed to a human source. Recognition accuracy for LLM feedback was only slightly above chance. In contrast, uptake of feedback was associated with its content, coded feedback quality, and the instructional composite, showing no association with the perceived or actual source. These findings underscore the disconnect between perception and uptake and emphasise the importance of addressing teachers' perceptions and biases regarding feedback sources.

## Data Availability Statement

The data that support the findings of this study, comprising the message-level quality coding, the participant-level responses, and an analysis script that reproduces all models reported in the article and in the supplementary material, are openly available in the Open Science Framework at [URL will be made available before publication].

## References

Aleven, V., McLaren, B. M., Sewall, J., van Velsen, M., Popescu, O., Demi, S., Ringenberg, M., & Koedinger, K. R. (2016). Example-tracing tutors: Intelligent tutor development for non-programmers. *International Journal of Artificial Intelligence in Education*, *26*(1), 224–269. https://doi.org/10.1007/s40593-015-0088-2

Atzmüller, C., & Steiner, P. M. (2010). Experimental vignette studies in survey research. *Methodology*, *6*(3), 128–138. https://doi.org/10.1027/1614-2241/a000014

Boud, D., & Molloy, E. (2013). Rethinking models of feedback for learning: The challenge of design. *Assessment & Evaluation in Higher Education*, *38(6)*, 698–712. https://doi.org/10.1080/02602938.2012.691462

Carless, D. (2006). Differing perceptions in the feedback process. *Studies in Higher Education*, *31*, 219–233. https://doi.org/10.1080/03075070600572132

Carless, D. (2015). *Excellence in university assessment: Learning from award-winning practice* (1st ed.). Routledge. https://doi.org/10.4324/9781315740621

Carless, D., & Boud, D. (2018). The development of student feedback literacy: Enabling uptake of feedback. *Assessment & Evaluation in Higher Education*, *43*(8), 1315–1325. https://doi.org/10.1080/02602938.2018.1463354

Chiu, T. K., Xia, Q., Zhou, X., Chai, C. S., & Cheng, M. (2023). Systematic literature review on opportunities, challenges, and future research recommendations of artificial intelligence in education. *Computers and Education: Artificial Intelligence*, *4*, 100118. https://doi.org/10.1016/j.caeai.2022.100118

Dai, W., Tsai, Y.S., Lin, J., Aldino, A., Jin, H., Li, T., Gašević, D., & Chen, G. (2024). Assessing the proficiency of large language models in automatic feedback generation: An evaluation study. *Computers and Education: Artificial Intelligence*, *7*. https://doi.org/10.1016/j.caeai.2024.100299

Demszky, D., Liu, J., Hill, H. C., Jurafsky, D., & Piech, C. (2023). Can automated feedback improve teachers' uptake of student ideas? Evidence from a randomized controlled trial in a large-scale online course. *Educational Evaluation and Policy Analysis*, *46*(3), 483-505. https://doi.org/10.3102/01623737231169270

Dijks, M. A., Brummer, L., & Kostons, D. (2018). The anonymous reviewer: the relationship between perceived expertise and the perceptions of peer feedback in higher education. *Assessment & Evaluation in Higher Education*, *43*(8), 1258–1271. https://doi.org/10.1080/02602938.2018.1447645

Er, E., Akçapınar, G., Bayazıt, A., Noroozi, O., & Banihashem, S. K. (2024). Assessing student perceptions and use of instructor versus AI -generated feedback. *British Journal of Educational Technology,* Advance online publication. https://doi.org/10.1111/bjet.13558

Ericsson, K. A., Krampe, R. T., & Tesch-Römer, C. (1993). The role of deliberate practice in the acquisition of expert performance. *Psychological Review, 100*(3), 363–406. https://doi.org/10.1037/0033-295X.100.3.363

Escalante, J., Pack, A., & Barrett, A. (2023). AI-generated feedback on writing: Insights into efficacy and ENL student preference. *International Journal of Educational Technology in Higher Education, 20*, 57. https://doi.org/10.1186/s41239-023-00425-2

Fleckenstein, J., Liebenow, L. W., & Meyer, J. (2023). Automated feedback and writing: A multi-level meta-analysis of effects on students' performance. *Frontiers in Artificial Intelligence*, *6*, 1162454. https://doi.org/10.3389/frai.2023.1162454

Fleckenstein, J., Meyer, J., Jansen, T., Keller, S. D., Köller, O., & Möller, J. (2024). Do teachers spot AI? Evaluating the detectability of AI-generated texts among student essays. *Computers and Education: Artificial Intelligence*, *6*. https://doi.org/10.1016/j.caeai.2024.100209

Fredricks, J. A., Blumenfeld, P. C., & Paris, A. H. (2004). School engagement: Potential of the concept, state of the evidence. *Review of Educational Research*, *74*(1), 59–109. https://doi.org/10.3102/00346543074001059

Gielen, M., & De Wever, B. (2015). Structuring peer assessment: Comparing the impact of the degree of structure on peer feedback content. *Computers in Human Behavior, 52*, 315–325. https://doi.org/10.1016/j.chb.2015.06.019

Gielen, S., Peeters, E., Dochy, F., Onghena, P., & Struyven, K. (2010). Improving the effectiveness of peer feedback for learning. *Learning and Instruction, 20*(4), 304–315. https://doi.org/10.1016/j.learninstruc.2009.08.007

Gigerenzer, G., & Gaissmaier, W. (2011). Heuristic decision making. Annual Review of Psychology, 62, 451–482. https://doi.org/10.1146/annurev-psych-120709-145346

Go, E., Jung, E. H., & Wu, M. (2014). The effects of source cues on online news perception. Computers in Human Behavior, 38, 358–367. https://doi.org/10.1016/j.chb.2014.05.044

Griffiths, C. M., Murdock-Perriera, L., & Eberhardt, J. L. (2023). "Can you tell me more about this?": Agentic written feedback, teacher expectations, and student learning. Contemporary Educational Psychology, 73, Article 102145. https://doi.org/10.1016/j.cedpsych.2022.102145

Hattie, J. (2023). Visible learning: The sequel: A synthesis of over 2,100 meta-analyses relating to achievement (1st ed.). Routledge. https://doi.org/10.4324/9781003380542

Hattie, J., & Timperley, H. (2007). The power of feedback. *Review of Educational Research*, *77*, 81–112. https://doi.org/10.3102/003465430298487

Hawkins, B., Taylor-Griffiths, D., & Lodge, J. M. (2025). Summarise, elaborate, try again: exploring the effect of feedback literacy on AI-enhanced essay writing. *Assessment & Evaluation in Higher Education*, 1–13. https://doi.org/10.1080/02602938.2025.2492070

Henderson, M., Ajjawi, R., Boud, D., & Molloy, E. (Eds.). (2019). The impact of feedback in higher education: Improving assessment outcomes for learners. Springer International Publishing. https://doi.org/10.1007/978-3-030-25112-3

Henderson, M., Bearman, M., Chung, J., Fawns, T., Buckingham Shum, S., Matthews, K. E., & de Mello Heredia, J. (2025). Comparing generative AI and teacher feedback: Student perceptions of usefulness and trustworthiness. *Assessment & Evaluation in Higher Education*, 1–16. https://doi.org/10.1080/02602938.2025.2502582

Hovland, C.I., Janis, I.L., & Kelley, H.H. (1953). Communication and persuasion. Yale University Press.

Ilgen, D. R., Fisher, C. D., & Taylor, M. S. (1979). Consequences of individual feedback on behavior in organizations. *Journal of Applied Psychology*, *64*(4), 349–371.

Jacobsen, L. J., Pargmann, J., Rohlmann, J., & Weber, K. E. (2026). Empirische Arbeit: KI-Literacy von Lehrenden: Prompt Engineering und Modellwahl als Prädiktoren der Qualität von KI-Feedback [AI literacy of educators: Prompt engineering and model choice as predictors of AI feedback quality]. *Psychologie in Erziehung und Unterricht, 73*(3), 137–152. https://doi.org/10.2378/peu2026.art10d

Jacobsen, L. J., & Weber, K. E. (2025). The promises and pitfalls of large language models as feedback providers: A study of prompt engineering and the quality of AI-driven feedback. AI, 6(2), 35. https://doi.org/10.3390/ai6020035

Jakesch, M., Hancock, J. T., & Naaman, M. (2023). Human heuristics for AI-generated language are flawed. *Proceedings of the National Academy of Sciences*, *120*(11), e2208839120. https://doi.org/10.1073/pnas.2208839120

Jansen, T., Höft, L., Bahr, L., Fleckenstein, J., Möller, J., Köller, O., & Meyer, J. (2024). Empirische Arbeit: Comparing Generative AI and Expert Feedback to Students' Writing: Insights from Student Teachers. Psychologie in Erziehung Und Unterricht, 71(2), 80–92. https://doi.org/10.2378/peu2024.art08d

Jansen, T., Horbach, A., & Meyer, J. (2025). Feedback from generative AI: Correlates of student engagement in text revision from 655 classes from primary and secondary school. In Proceedings of the 2025 ACM Conference (pp. 831–836). Association for Computing Machinery. https://doi.org/10.1145/3706468.3706494

Joughin, G., Dawson, P., & Boud, D. (2017). Improving assessment tasks through addressing our unconscious limits to change. Assessment & Evaluation in Higher Education, 42(8), 1221–1232. https://doi.org/10.1080/02602938.2016.1257689

Kahneman, D. (2011). Thinking, fast and slow. Penguin.

Kahneman, D., & Tversky, A. (1979). Prospect theory: An analysis of decision under risk. Econometrica, 47(2), 263–292. https://www.jstor.org/stable/1914185

Kao, C. W., & Reynolds, B. L. (2024). Timed second language writing performance: Effects of perceived teacher vs perceived automated feedback. *Humanities and Social Sciences Communications*, *11*, Article 1012. https://doi.org/10.1057/s41599-024-03522-3

Kluger, A. N., & DeNisi, A. (1996). The effects of feedback interventions on performance: A historical review, a meta-analysis and a preliminary feedback intervention theory. *Psychological Bulletin*, *119*(2), 254–284. https://doi.org/10.1037/0033-2909.119.2.254

Kraft, M. A., Blazar, D., & Hogan, D. (2018). The effect of teacher coaching on instruction and achievement: A meta-analysis of the causal evidence. *Review of Educational Research, 88*(4), 547–588. https://doi.org/10.3102/0034654318759268

Kulik, J. A., & Fletcher, J. D. (2016). Effectiveness of intelligent tutoring systems: A meta-analytic review. *Review of Educational Research, 86*(1), 42–78. https://doi.org/10.3102/0034654315581420

Lipnevich, A. A., Berg, D. A., & Smith, J. K. (2016). Toward a model of student response to feedback. *Handbook of human and social conditions in assessment* (pp. 169-185). Routledge.

Lipnevich, A. A., & Panadero, E. (2021). A review of feedback models and theories: Descriptions, definitions, and conclusions. Frontiers in Education, 6, Article 720195. https://doi.org/10.3389/feduc.2021.720195

Lipnevich, A. A., & Smith, J. K. (2022). Student–Feedback Interaction Model: Revised. *Studies in Educational Evaluation, 75*, Article 101208. https://doi.org/10.1016/j.stueduc.2022.101208

Lu, H.-L. (2010). Research on peer-coaching in preservice teacher education: A review of literature. *Teaching and Teacher Education, 26*(4), 748–753. https://doi.org/10.1016/j.tate.2009.10.015

Meinert, J., & Krämer, N. C. (2022). How the expertise heuristic accelerates decision-making and credibility judgments in social media by means of effort reduction. PLOS ONE, 17(3), e0264428. https://doi.org/10.1371/journal.pone.0264428

Mertens, U., Jansen, T., Steinbach, M., Fleckenstein, J., & Meyer, J. (2025). Perceptions of teacher- versus AI-generated feedback: Experimental findings on the (implicit) bias of teachers against AI [Preprint]. PsyArXiv. https://doi.org/10.31234/osf.io/s25rn

Meyer, J., Jansen, T., Schiller, R., Liebenow, L. W., Steinbach, M., Horbach, A., & Fleckenstein, J. (2024). Using LLMs to bring evidence-based feedback into the classroom: AI-generated feedback increases secondary students' text revision, motivation, and positive emotions. Computers and Education: Artificial Intelligence, 6, 100199. https://doi.org/10.1016/j.caeai.2023.100199

Moorhouse, B. L., & Kohnke, L. (2024). The effects of generative AI on initial language teacher education: The perceptions of teacher educators. *System, 122*, 103290. https://doi.org/10.1016/j.system.2024.103290

Narciss, S. (2008). Feedback strategies for interactive learning tasks. In J. M. Spector, M. D. Merrill, J. J. G. Van Merriënboer, & M. P. Driscoll (Eds.), *Handbook of research on educational communications and technology* (3rd ed., pp. 125–143). Erlbaum.

Narciss, S. (2013). Designing and evaluating tutoring feedback strategies for digital learning environments on the basis of the interactive feedback model. *Digital Education Review*, *23*(1), 7–26. https://doi.org/10.1344/der.2013.23.7-26

Nazaretsky, T., Mejia-Domenzain, P., Swamy, V., Frej, J., & Käser, T. (2024). AI or human? Evaluating student feedback perceptions in higher education. In R. F. Mello, N. Rummel, I. Jivet, G. Pishtari, & J. A. Ruipérez Valiente (Eds.), *Technology Enhanced Learning for Inclusive and Equitable Quality Education* (pp. 284–298). Springer Nature Switzerland. https://doi.org/10.31219/osf.io/6zm83

Panadero, E. (2023). Toward a paradigm shift in feedback research: Five further steps influenced by self-regulated learning theory. *Educational Psychologist*, *58*(3), 193–204. https://doi.org/10.1080/00461520.2023.2223642

Panadero, E., & Lipnevich, A. A. (2022). A review of feedback models and typologies: Towards an integrative model of feedback elements. *Educational Research Review*, *35*, Article 100416. https://doi.org/10.1016/j.edurev.2021.100416

Pataranutaporn, P., Liu, R., Finn, E., Avila, E., Chien, A., Maes, P., & McDuff, D. (2023). Influencing human–AI interaction by priming beliefs about AI can increase perceived trustworthiness, empathy and effectiveness. *Nature Machine Intelligence*, *5*, 1076–1086. https://doi.org/10.1038/s42256-023-00720-7

Prilop, C. N., Mah, D., Jacobsen, L. J., Hansen, R. R., Weber, K. E., & Hoya, F. (2024). Generative AI in teacher education: Using AI-enhanced methods to explore teacher educators' perceptions. https://doi.org/10.31219/osf.io/szcwb

Prilop, C. N. & Weber, K. E. (2023). Digital video-based peer feedback training: the effect of expert feedback on pre-service teachers' peer feedback beliefs and peer feedback quality. Teaching and Teacher Education, 104099. https://doi.org/10.1016/j.tate.2023.104099

Prilop, C. N., Weber, K. E., & Kleinknecht, M. (2019). Entwicklung eines video- und textbasierten Instruments zur Messung kollegialer Feedbackkompetenz von Lehrkräften. In Lehrer. Bildung. Gestalten.: Beiträge zur empirischen Forschung in der Lehrerbildung. Beltz Juventa.

Prilop, C. N., Weber, K. E., & Kleinknecht, M. (2020). Effects of digital video-based feedback environments on pre-service teachers' feedback competence. Computers in Human Behavior, 102, 120–131. https://doi.org/10.1016/j.chb.2019.08.011

Prilop, C. N., Weber, K. E., Prins, F. & Kleinknecht, M. (2021). Connecting feedback to self-efficacy: Receiving and providing peer feedback in teacher education. Studies in Educational Evaluation, 70(2), 101062. https://doi.org/10.1016/j.stueduc.2021.101062

Prins, F., Sluijsmans, D., & Kirschner, P. A. (2006). Feedback for general practitioners in training: Quality, styles and preferences. *Advances in Health Sciences Education*, *11*, 289–303. https://doi.org/10.1007/s10459-005-3250-z

Qin, J., & Uccelli, P. (2022). Automated writing evaluation feedback for writing: A systematic review. *Computer Assisted Language Learning*, *35*(8), 1-29. https://doi.org/10.1080/09588221.2022.2033787

Razzaq, L., Patvarczki, J., Almeida, S. F., Vartak, M., Feng, M., Heffernan, N. T., & Koedinger, K. R. (2009). The ASSISTment Builder: Supporting the life cycle of tutoring system content creation. *IEEE Transactions on Learning Technologies, 2*(2), 157–166. https://doi.org/10.1109/TLT.2009.23

Robinson, M. D., & Clore, G. L. (2001). Simulation, scenarios, and emotional appraisal: Testing the convergence of real and imagined reactions to emotional stimuli. *Personality and Social Psychology Bulletin*, *27*(11), 1520–1532. https://doi.org/10.1177/01461672012711012

Rubin, M., Li, J. Z., Zimmerman, F., Ong, D. C., Goldenberg, A., & Perry, A. (2025). Comparing the value of perceived human versus AI-generated empathy. *Nature Human Behaviour*. Advance online publication. https://doi.org/10.1038/s41562-025-02247-w

Ruwe, T., & Kuklick, L. (2025). Quality counts? Examining the role of feedback provider and feedback quality on students' feedback perceptions. *British Journal of Educational Technology*. Advance online publication. https://doi.org/10.1111/bjet.70011

Ruwe, T., & Mayweg-Paus, E. (2024). Embracing LLM feedback: The role of feedback providers and provider information for feedback effectiveness. *Frontiers in Education*, *9*. https://doi.org/10.3389/feduc.2024.1461362

Smith, J. K., & Lipnevich, A. A. (2018). Instructional feedback: Analysis, synthesis, and extrapolation. In A. A. Lipnevich & J. K. Smith (Eds.), *The Cambridge handbook of instructional feedback* (Cambridge handbooks in psychology, pp. 591–604). Cambridge University Press. https://doi.org/10.1017/9781316832134.021

Stanovich, K. E., & West, R. F. (2000). Individual differences in reasoning: Implications for the rationality debate? Behavioral and Brain Sciences, 23(5), 645–665. https://doi.org/10.1017/S0140525X00003435

Steenbergen-Hu, S., & Cooper, H. (2014). A meta-analysis of the effectiveness of intelligent tutoring systems on college students' academic learning. *Journal of Educational Psychology, 106*(2), 331–347. https://doi.org/10.1037/a0034752

Steiss, J., Tate, T., Graham, S., Cruz, J., Hebert, M., Wang, J., Moon, Y., Tseng, W., Warschauer, M., & Olson, C. B. (2024). Comparing the quality of human and ChatGPT feedback of students' writing. *Learning and Instruction*, *91*, 101894. https://doi.org/10.1016/j.learninstruc.2024.101894

Strijbos, J.-W., Narciss, S., & Dünnebier, K. (2010). Peer feedback content and sender's competence level in academic writing revision tasks: Are they critical for feedback message perceptions and efficiency? *Learning and Instruction*, *20(4)*, 291–303. https://doi.org/10.1016/j.learninstruc.2009.08.008

Strijbos, J.-W., Pat-El, R., & Narciss, S. (2021). Structural validity and invariance of the Feedback Perceptions Questionnaire. *Studies in Educational Evaluation*, *68*, 100980. https://doi.org/10.1016/j.stueduc.2021.100980

Tajfel, H., & Turner, J. C. (1979). An integrative theory of intergroup conflict. In W. G. Austin & S. Worchel (Eds.), The social psychology of intergroup relations (pp. 33–47). Brooks/Cole.

Tversky, A., & Kahneman, D. (1989). Rational choice and the framing of decisions. In B. Karpak & S. Zionts (Eds.), Multiple criteria decision making and risk analysis using microcomputers (NATO ASI Series, Vol. 56, pp. 81–126). Springer. https://doi.org/10.1007/978-3-642-74919-3_4

UNESCO. (2024). AI competency framework for teachers. United Nations Educational, Scientific and Cultural Organization. https://discovery.ucl.ac.uk/id/eprint/10196729/1/UNESCO_AI_CFT_Final.pdf

Van der Kleij, F. M., & Lipnevich, A. A. (2021). Student perceptions of assessment feedback: A critical scoping review and call for research. *Educational Assessment, Evaluation and Accountability, 33*(2), 345–373. https://doi.org/10.1007/s11092-020-09331-x

Weber, K. E., Gold, B., Prilop, C. N., & Kleinknecht, M. (2018). Promoting pre-service teachers' professional vision of classroom management during practical school training: Effects of a structured online- and video-based self-reflection and feedback intervention. Teaching and Teacher Education, 76, 39–49. https://doi.org/10.1016/j.tate.2018.08.008

Weidlich, J., Gotsch, F., Schudel, K., Marusic-Würscher, C., Bolten, H., Luger, S., Bütler, D., Wohlfehnder, B., Mazzarella, J., & Maag Merki, K. (2026). The medium is not the message: Message features in teacher, peer, and AI feedback help explain the divergence of perceptions and performance. *SocArXiv*. https://doi.org/10.31235/osf.io/t5m4k_v1

Winstone, N. E., Nash, R. A., Parker, M., & Rowntree, J. (2017). Supporting learners' agentic engagement with feedback: A systematic review and a taxonomy of recipience processes. *Educational Psychologist*, *52*(1), 17–37. https://doi.org/10.1080/00461520.2016.1207538

Wisniewski, B., Zierer, K., & Hattie, J. (2020). The power of feedback revisited: A meta-analysis of educational feedback research. *Frontiers in Psychology*, *10*, Article 3087. https://doi.org/10.3389/fpsyg.2019.03087

Yin, B., Li, J., Gao, M., Li, Y., & Xu, B. (2024). Using a chatbot to provide formative feedback: A longitudinal study of intrinsic motivation, cognitive load, and learning performance. *Behavioral Sciences*, *15*(5), 152. https://doi.org/10.3390/bs15050152

Zhang, A., Gao, Y., Suraworachet, W., Nazaretsky, T., & Cukurova, M. (2025). Evaluating trust in AI, human, and co-produced feedback among undergraduate students, *arXiv*. https://doi.org/10.48550/arXiv.2504.10961

Ziegler, R., von Schwichow, A., & Diehl, M. (2005). Matching the message source to attitude functions: Implications for biased processing. Journal of Experimental Social Psychology, 41, 645–653. https://doi.org/10.1016/j.jesp.2004.12.002